\documentclass[english]{IEEEtran}
\usepackage[T1]{fontenc}
\usepackage[latin9]{inputenc}
\usepackage{array}
\usepackage{textcomp}
\usepackage{multirow}
\usepackage{amstext}
\usepackage{graphicx}

\makeatletter

\providecommand{\tabularnewline}{\\}

\author{I.~Badillo
J.~Jugo,
J.~Portilla,
J.~Feutchwanger,
C.~San~Vicente,
V.~Etxebarria
\thanks{Department of Electricity and Electronics, University of the Basque Country UPV/EHU, Leioa, 48940, Bizkaia, SPAIN, e-mail: inari.badillo@ehu.es}
}

\makeatother

\usepackage{babel}
\begin{document}

\title{PXIe-based LLRF architecture and versatile test bench for heavy ion
linear acceleration}
\maketitle
\begin{abstract}
This work describes the architecture of a digital LLRF system for
heavy-ion acceleration developed under the specification of the projected
future heavy-ion accelerator facility in Huelva, Spain. A prototype
LLRF test bench operating at 80MHz in CW mode has been designed and
built. The core LLRF control has been digitally implemented on a PXIe
chassis, including an FPGA for digital signal processing and a real
time controller. The test bench is completed with a good quality signal
generator used as master frequency reference, an analog front end
for reference modulation and signal conditioning, small RF components
completing the circuit, as well as a tunable resonant cavity at 80
MHz, whose RF amplitude, phase and frequency are real-time controlled
and monitored. The presented LLRF system is mainly digitally implemented
using a PXIe platform provided by National Instruments, and is based
on IQ modulation and demodulation. The system can be configured to
use both direct sampling and undersampling techniques, resulting thus
in a high performance and versatile RF control system without the
need of excessive computational resources or very high speed acquisition
hardware. All the system is programmed using the LabVIEW environment,
which makes much easier the prototyping process and its reconfigurability. \end{abstract}
\begin{IEEEkeywords}
LLRF, undersampling, IQ modulation, fast DAQ, PXI/PXIe, LabVIEWº
\end{IEEEkeywords}

\section{Introduction}

Low-level Radio Frequency (LLRF) control systems are essential parts
of modern particle accelerators. Their main task is transferring energy
to the beam in a controlled fashion by properly governing the RF accelerating
fields and synchronizing them to the particle bunches \cite{cern}.

A typical LLRF control system usually consists of a fast loop to
regulate the amplitude and the phase of the accelerating voltage as
seen by the particles, and one slower loop to tune the resonant frequency
of the accelerating cavities. The amplitude and phase loop should
have a wide bandwidth, up to some hundred kHz for typical accelerators,
and should be able to compensate the ripples of the high voltage power
sources and other perturbations as well as to have a good time response,
particularly when pulsed RF fields are required. The frequency loop,
which has a much lower bandwidth, usually in the range of a few hundred
Hz, controls the cavity tuners to maintain its nominal resonance frequency
in the presence of cavity warming, mechanical perturbations and beam
loading, so that the reflected power is minimized \cite{CSNS/RCS}.

LLRF systems are typically implemented using analog or digital electronics
or a combination of both \cite{digAnalog1,digAnalog2,digAnalog3}.
The analog approach is comparatively cheap and provides high bandwidth
and short delays, very well fitting the usual RF control requirements.
However, modern digital solutions also fulfill these requirements
while providing also much higher flexibility, reprogrammability and
stability of the controlled variables \cite{digLLRF,LLRFdesign}.
Depending on the specifications of the particular operation facility,
LLRF systems are usually conceived and implemented differently, and
generic reconfigurable solutions, suitable for a number of different
acceleration facilities are not commonly developed \cite{character}.

Being the LLRF system one of the key elements in a particle accelerator,
the design of such system is an active working area, requiring important
developing effort.

This work describes the architecture of a new digital LLRF system
for heavy-ion acceleration developed under the specification of the
projected future accelerator facility in Huelva, Spain\cite{huelva}.
The presented system features two main ingredients which make it quite
unique. On the one hand the proposed digital acquisition and processing
architecture makes use of state-of-the-art undersampling techniques
\cite{undersampling}, leading to a high performance RF control system
without the need of excessive computational resources or expensive
high speed acquisition hardware. On the other hand the presented solution
is conceived to be highly modular and reconfigurable while it makes
use of widespread PXIe hardware, well maintained by the industry,
which combines high performance, scalability and high availability,
meaning that the LLRF design proposed here can be easily adapted to
many different situations, specifications and facilities. 

As experimental evaluation of the presented architecture, a prototype
LLRF test bench operating at $80MHz$ in CW mode has been designed,
built and tested in the laboratory. The core LLRF control has been
digitally implemented on a PXIe chassis, including an FPGA for digital
signal processing and real time control. The signal acquisition is
performed in digitizers following the FlexRIO reconfigurable I/O (RIO)
architecture, by direct sampling or undersampling, with real time
communication between the processing boards with peer-to-peer FIFO
channels. All the system is programmed using the LabVIEW environment,
which makes much easier the prototyping process and its reconfigurability.
Moreover, LabVIEW is also used for the real time monitoring of the
signals and the adjustment of the system parameters. Thus, the system
is a very valuable tool for testing purposes of new techniques and
ideas.

The paper is structured as follows: In next section a brief description
of the signal processing and undersampling techniques used for our
LLRF system is given. The third section is devoted to describing the
laboratory experimental setup where the proposed modular and reconfigurable
RF control system has been tested. In the following section, the obtained
experimental results are presented and finally some concluding remarks
are included in last section.

\section{Signal acquisition and processing system}

The reference RF signal used to feed the current test bench is obtained
from an RF generator. It is a $80MHz$ signal, which is the nominal
resonant frequency of the cavity. This frequency leads to fast sampling
rates using a direct digitalization scheme without the implementation
of an IF mixing technique if standard sampling techniques (oversampling)
are used, which means higher equipment costs. The sampling technique
known as undersampling or subsampling allows to use lower sampling
rates without losing information. In this section, the subsampling
technique is outlined, describing the hardware selected for data acquisition
and processing.

\subsection{Subsampling technique}

The Nyquist sampling theorem is a well known theoretical result, which
is widely used in signal processing and control applications. This
theorem states that for a band limited signal, with a limit $\omega_{0}$
, the sequence obtained sampling this signal with a sampling frequency
$\omega_{s}$ fulfilling 
\begin{equation}
\omega_{s}>2\omega_{0}\label{eq: Nyquist}
\end{equation}
has the same information as the original continuous signal and the
aliasing phenomenon will not lead to ambiguous data. The limit of
the previous condition defines the Nyquist rate as the minimum sampling
rate to avoid aliasing. The main disadvantage of this technique is
the necessity of high sampling frequencies. 

However, a modified version of this theorem, known as Nyquist-Shannon
sampling theorem, says that the sampling frequency needs to be twice
the signal bandwidth and not twice the maximum frequency component,
in order to be able to reconstruct the original signal perfectly,
under several conditions, from the sampled version. That is,

\begin{equation}
\omega_{s}>2BW\label{eq:Nyquist-Shannon-1}
\end{equation}
In this case, the aliasing at lower frequencies can appear but without
leading to overlap of frequency components. Then, under condition
\ref{eq:Nyquist-Shannon-1}, the undersampling or bandpass sampling,
that is, the use of sampling frequencies not fulfilling condition
\ref{eq: Nyquist}, will lead to unambiguous data, \cite{undersampling}.
The main advantage is that the sampling frequency is directly related
with the signal bandwidth and not to the higher frequency component.
However, the design of an application based on undersampling must
consider carefully the bandwidth of the continuous signals and the
filtering necessities.

\begin{figure}
\centering{}\includegraphics[width=0.9\columnwidth]{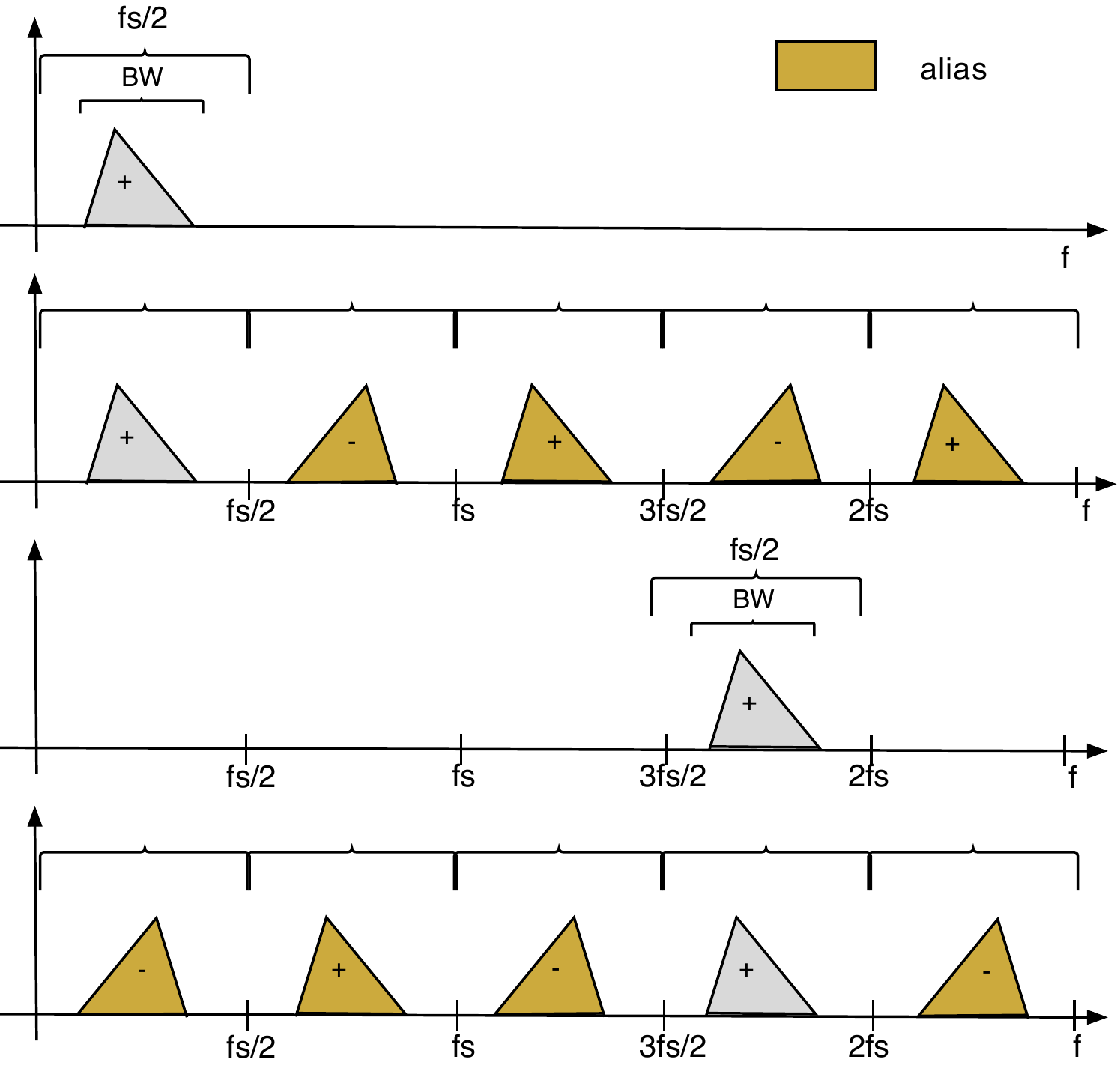}\protect\caption{\label{fig:Difference-between-oversampling}Difference between oversampling
(up) and undersampling (down) of a band limited (BW) signal with sampling
frequency $f_{s}$. In this case, the spectral inversion case can
be observed}
\end{figure}

The differences between the use of oversamping and undersampling can
be observed in Figure \ref{fig:Difference-between-oversampling}.

One important thing to consider is the effect of the harmonics of
the original signal, since a real ADC will have nonlinear effects
adding high order harmonics, leading to the existence of Nyquist aliasing
in an analog to digital converter due to those high harmonics \cite{subnyquist}.
For this reason, the selection of an adequate sampling frequency is
fundamental for determining the location of the best frequency space
in an analog to digital converter to avoid excessive Nyquist aliasing
of high order harmonics. Then, the sampling frequency should be chosen
in the next range to avoid aliasing: 
\begin{equation}
\frac{2f_{center}-BW}{m}\geq f_{s}\geq\frac{2f_{center}+BW}{m+1}\label{eq:rangos}
\end{equation}
where $m$ is an integer and $f_{center}$ is the central frequency
of the band limited signal to sample.

Note that by applying undersampling, the replication of the spectral
components of the signals can lead to a spectral inversion, depending
on the election of the sampling signal (for instance, see Figure \ref{fig:Difference-between-oversampling}).
To avoid this phenomenon, the sampling frequency should fulfill the
next equation:

\textbf{
\begin{equation}
f_{s}=\frac{2f_{center}-BW}{m_{even}}\label{eq:impar}
\end{equation}
}where $m_{even}=2,4,6,\dots$ is a even positive integer.

An additional practical condition is the selection of the sampling
frequency to be centered in the sampled spectra at $\pm f_{s}/4$,
which simplifies several operations such as filtering. In such case,
to ensure this condition, the sampling frequency is selected using
the next equations:

\textbf{
\begin{equation}
f_{s}>2BW\qquad f_{s}=\frac{4f_{center}}{2m-1}\label{eq:condition}
\end{equation}
}where $m$ is a positive integer.

This technique can be applied to sample high speed signals with lower
sampling frequencies, as for example RF signals in a particle accelerator,
by choosing carefully the sampling frequency, on the one hand, and
by filtering the signals to avoid distortion due to undesired components
without information out of the band of interest, on the other hand.
In the next section, a LLRF test bench using direct sampling of RF
signals, based on undersampling is presented.

In the particular application shown in this work, a $80MHz$ sinewave
is used so $f_{center}=80MHz$ and BW can be assumed to be narrow,
depending on the Q cavity factor. In fact, the effective bandwidth
can be limited designing a convenient control system. So, for $m=3$,
the suitable sampling frequency range goes from $40MHz$ to $53,33MHz$,
not limiting the signal BW.

Therefore, the nominal sampling rate of the FlexRIO 5751R ADC (50MS/s)
is a valid sampling rate for our purposes, although some signal loss
is introduced using this card for a signal frequency of $80MHz$.
In this way, the alias centered at $\left(f_{0}+m\cdot f_{s}\right)$
will be used to reconstruct the original signal. In the particular
case of the present work this value is $80MHz-2\cdot50MHz=-20MHz$,
so the the whole information of the original signal will be extracted
from the acquired $20MHz$ signal. Note that the spectral inversion
observed in this case must be considered in the signal processing.

For a certain sampling point, the timing uncertainty (clock jitter
or clock phase noise) creates amplitude variation \cite{jitter}.
As the input frequency increases, a fixed amount of clock jitter leads
to a higher amplitude error. It is worth to point out that jitter
is does not cause an excessive error in measurements for this particular
application, since the original signal frequency and the sampling
rate are not far. 

In addition to undersampling-based data acquisition, conventional
sampling (oversampling) has been also performed for different tests
using a 250MS/s FlexRIO card. The combination of oversampling and
undersampling techniques makes the system very versatile.

\subsection{Data Acquisition and processing}

In order to acquire and monitor all the signals needed in the current
test bench, a digital real-time system has been developed, using a
solution based on the National Instruments PXIe architecture in addition
to the FlexRIO cards.

In the current implementation, the system consists of an 8 slot NI
1082 PXIe chassis with a PXIe 8108 RT embedded controller (dual core
at 2.53GHz CPU) running LabVIEW Real-Time operating system and two
different high throughput NI FlexRIO cards along with their respective
adapter modules: one for acquisition that includes an ADC up to 50MS/s
together with sixteen analog input channels, and the other one for
data generation which has two differential channels up to 1.25GS/s.
For slow signals, a multifunction DAQ card is also used under the
PXIe system, as well as a regular desktop PC to develop the code that
will be executed in both the PXIe controller and the FPGAs. More detailed
specifications of the PXI modules can be found in Table \ref{tab:Technical-specifications-of}.

\begin{table*}
\centering{}%
\begin{tabular}{|c|c|c|c|c|c|c|c|}
\hline 
 & Model & FPGA & DMA channels & Adapter module & ADC/DAC rate & Nº of channels & Resolution\tabularnewline
\hline 
\hline 
FlexRIO card 1 & NI PXIe-7961R & Virtex-5 SX50T & 16 & NI 5751 & 50MS/s & 16 (single-ended) & 14\tabularnewline
\hline 
FlexRIO card 2 & NI PXIe-7966R & Virtex-5 SX95T & 16 & AT-1212 & 1.25GS/s & 2 (differential) & 14\tabularnewline
\hline 
Multifunction card 1 & NI-7852R & Virtex-5 LX50 & 3 & - & 1 MS/s & 8AI, 8AO, 96DIO & 16\tabularnewline
\hline 
Multifunction card 2 & NI-PXI 6259 & - & - & - & \multirow{1}{*}{1MS/s} & 32 AI, 4AO, 48DIO & 16\tabularnewline
\hline 
\end{tabular}\protect\caption{\label{tab:Technical-specifications-of}Technical specifications of
the PXIe modules used in the developed digital system}
\end{table*}

This configuration gives a flexible solution, whose functionality
can be easily changed thanks to the modularity and reconfigurable
nature of the system.

\section{Test bench description }

\begin{figure*}
\begin{centering}
\includegraphics[width=0.8\paperwidth]{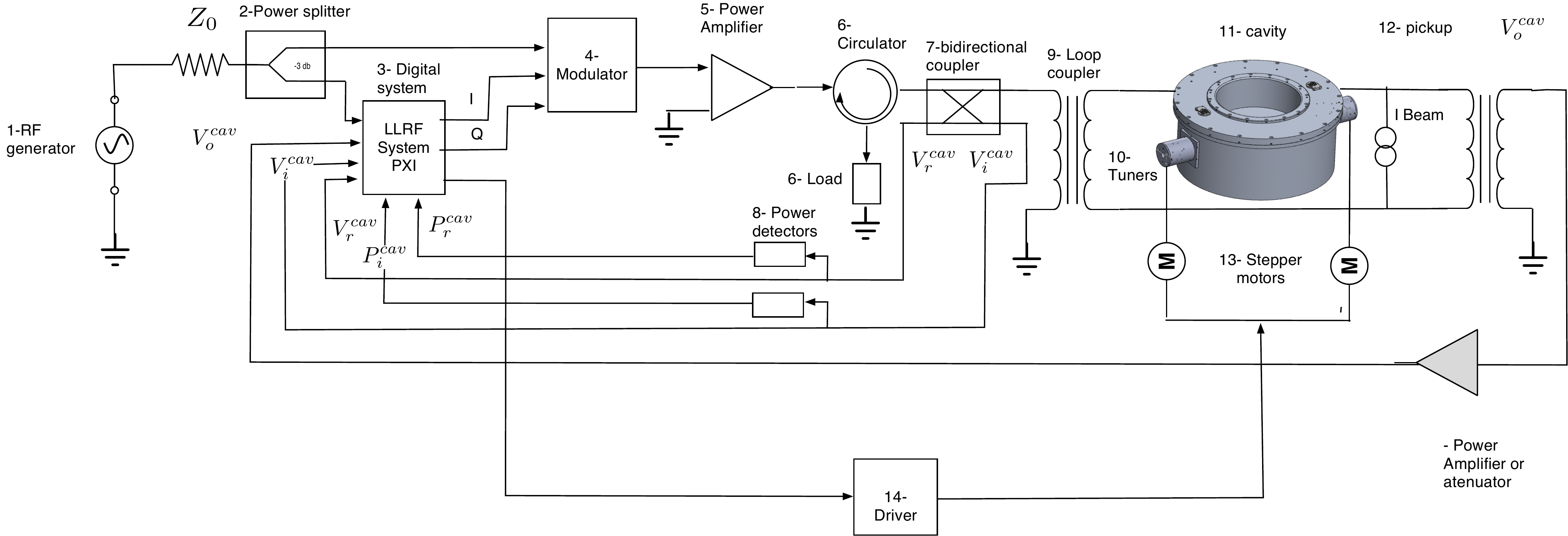}
\par\end{centering}

\centering{}\protect\caption{\label{fig:Experimental-setup-description}Basic description of the
Experimental setup }
\end{figure*}

The experimental setup is described in Figure \ref{fig:Experimental-setup-description}. 

\begin{figure}
\centering{}\includegraphics[width=0.9\columnwidth]{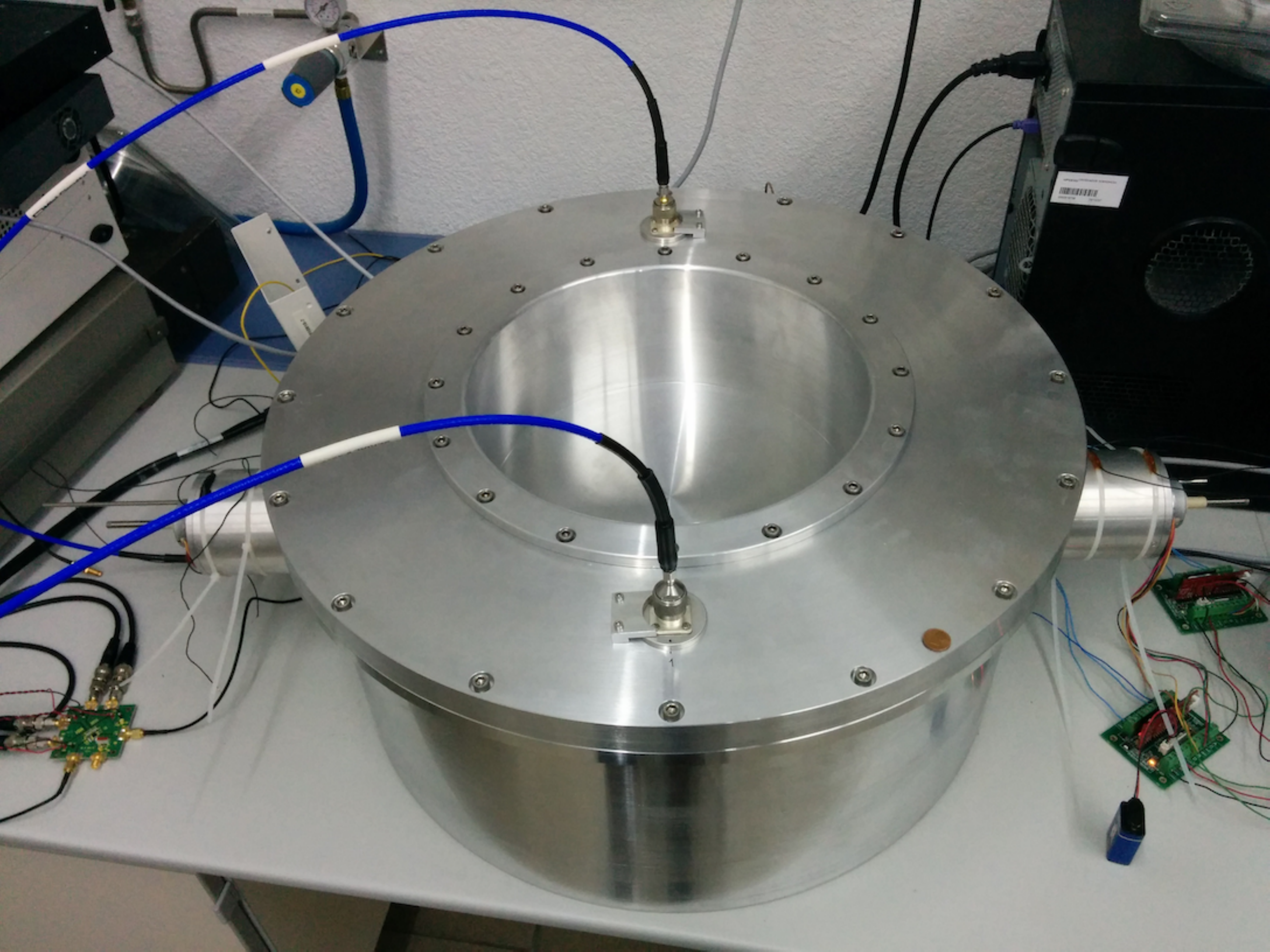}\protect\caption{The resonant cavity. \label{fig:The-resonant-cavity.}}
\end{figure}

The central element of the test bench is a tunable resonant cavity
at $80MHz$, see Figure \ref{fig:The-resonant-cavity.}. It is a re-entrant
type cavity machined in aluminum, with the electric field concentrated
on an accelerating gap near the axis, and the magnetic field mostly
located on the outer perimeter. The unloaded Q factor value of the
cavity has been designed to be around 6000, and two mechanical plungers
driven by stepper motors have been included in radial direction for
tuning the resonance frequency. Two rotatable loop coupling to the
magnetic field are used as RF input (drive) and output (pickup) couplers
in the cavity, respectively. 

\begin{figure}
\begin{centering}
\includegraphics[width=0.9\columnwidth]{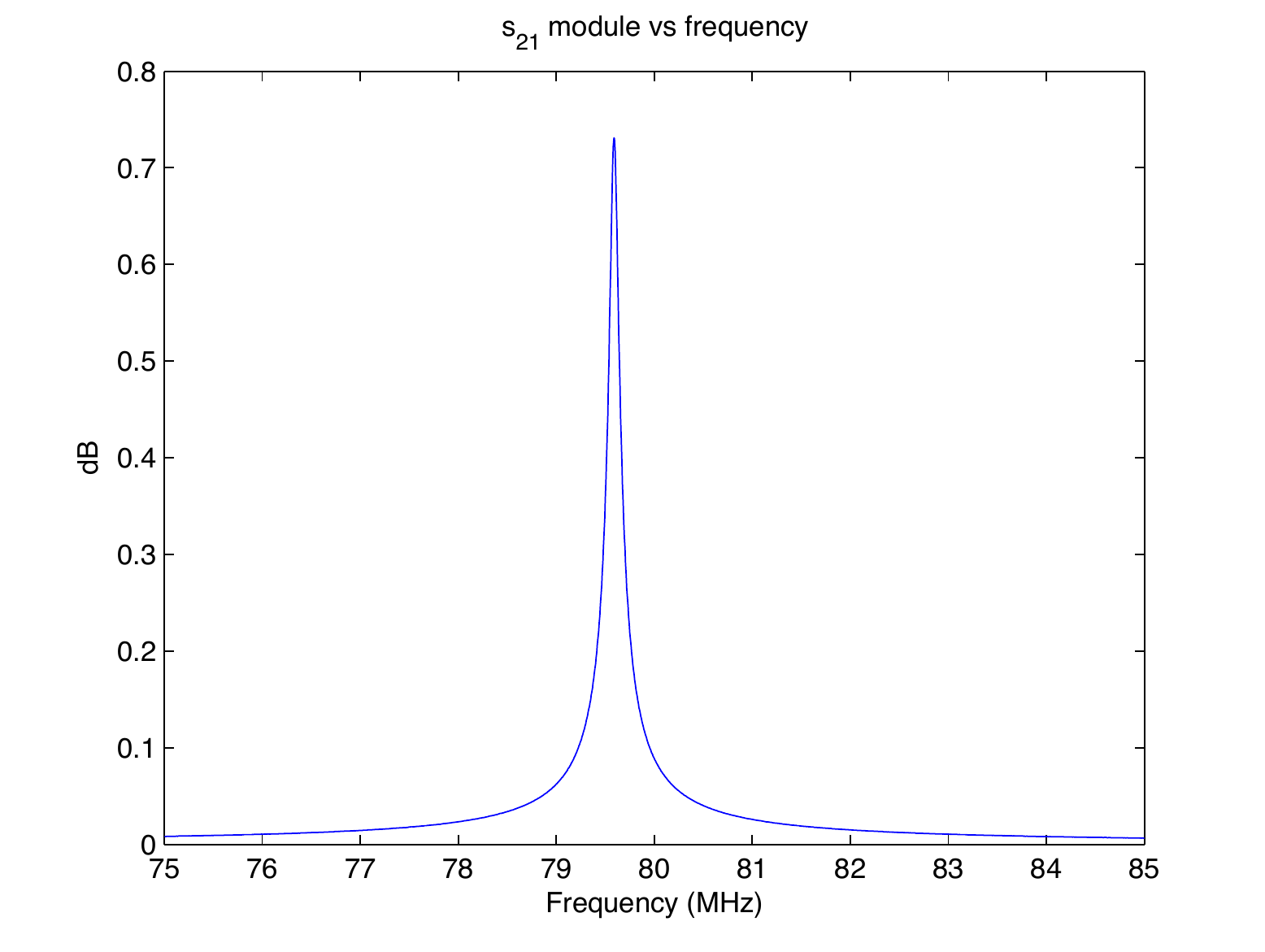}
\par\end{centering}

\centering{}\includegraphics[width=0.9\columnwidth]{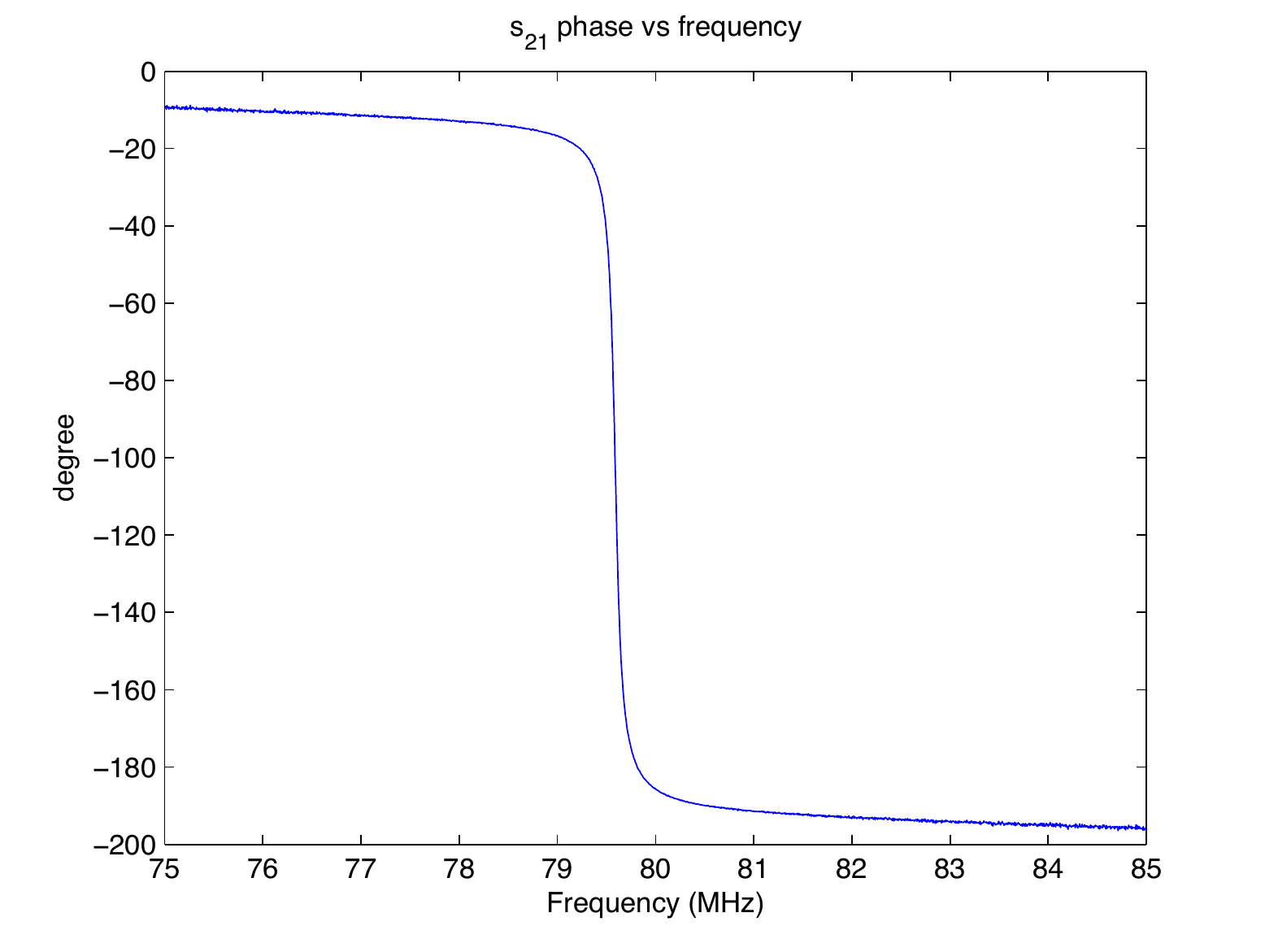}\protect\caption{\label{fig:Frequency-response-of}Frequency response of the cavity
in amplitude and phase}
\end{figure}

A characterization of the cavity showed that the resonant frequency
was slightly shifted from the nominal value, being located at $79,59MHz$
as shown in Figure \ref{fig:Frequency-response-of}. This deviation
does not cause any problem to the developed LLRF control. 

The developed LLRF system is digital, and has been conceived to be
a highly scalable, flexible and reconfigurable structure. So, it is
possible to introduce substantial changes in the design in order to
add or modify functionalities simply changing the program code. Another
advantage of digital LLRFs is that a higher amplitude and phase stability
can be achieved compared to analog setups \cite{Hooman}. The drawbacks
are lower speed, as well as higher cost and complexity.

The LLRF control is based on the PXIe architecture along with FPGA
technology. The core is a PXIe chassis running a RT OS in the controller
along with several acquisition and generation cards. The whole system
is under a LabVIEW environment. This is a good advantage since it
allows to easily integrate a wide catalog of hardware devices, reducing
the effort of complex tasks like peer-to-peer streaming or DMA transfers.
The use of LabVIEW also facilitates and speeds up the development
of monitorization and control structures thanks to the provided programming
tools. The possibility of programming FPGAs using LabVIEW code must
be highlighted, avoiding the use of long and complex HDL coding which
would prevent rapid prototyping.

As shown in Figure \ref{fig:Experimental-setup-description}, the
proposed LLRF system has six inputs and three outputs. These inputs
are: 
\begin{itemize}
\item The reference signal from RF generator.
\item $V_{o}^{cav}$, the cavity output from the pickup.
\item $V_{i}^{cav}$, Incident signal from a bidirectional coupler at the
cavity input.
\item $V_{r}^{cav}$, Reflected signal from a bidirectional coupler at the
cavity input.
\item $P_{i}^{cav}$, the power level of $V_{i}^{cav}$.
\item $P_{r}^{cav}$, the power level of $V_{r}^{cav}$.
\end{itemize}
And outputs:
\begin{itemize}
\item I, in-phase component.
\item Q, quadrature component.
\item Pulse train output for the stepper motors.
\end{itemize}
As mentioned before, two feedback control loops are implemented in
order to maintain the stability of the RF fields. The first one is
the phase and amplitude loop. The aim of this fast loop is to set
the RF gap voltage of the cavity and the phase as required, as well
as keeping both process variables stable within acceptable range.
In heavy ion LLRFs, these requirements are typically $<1\%$ in amplitude
and $<1\text{º}$ in phase error \cite{Hooman}. The main sources
for amplitude and phase perturbations are beam loading effects, variations
caused by nonlinearities in the amplifier and phase shifts caused
by temperature drifts in different devices \cite{DESY}.

This phase and amplitude loop is based on I/Q detection and control.
In this control scheme, in-phase and quadrature components of the
RF cavity field are calculated and controlled separately. 

A given sinusoidal signal $x(t)$ with a given A amplitude, $\omega$
frequency and $\varphi$ phase, can be represented into its in-phase
and quadrature components:

\begin{equation}
x(t)=Asin(\omega t+\varphi)=Acos(\varphi)sin(\omega t)+Asin(\varphi)cos(\omega t)\label{eq:sin}
\end{equation}

Where,

\begin{eqnarray}
I & = & Acos(\varphi)\label{eq:I}\\
Q & = & Asin(\varphi)\label{eq:Q}
\end{eqnarray}

This way, the amplitude and phase of the signal can be expressed in
terms of I and Q:

\begin{eqnarray}
A & = & \sqrt{I^{2}+Q^{2}}\label{eq:amplitud}\\
\varphi & = & atan(\frac{Q}{I})\label{eq:fase}
\end{eqnarray}

Converting the RF signal into I/Q components, allows to manipulate
the amplitude and phase without handling directly the carrier signal.
It is also advantageous because the symmetry of the I/Q signals \cite{iq}.
This scheme allows to control both phase and amplitude in a single
loop.

\begin{figure}
\centering{}\includegraphics[width=1\columnwidth]{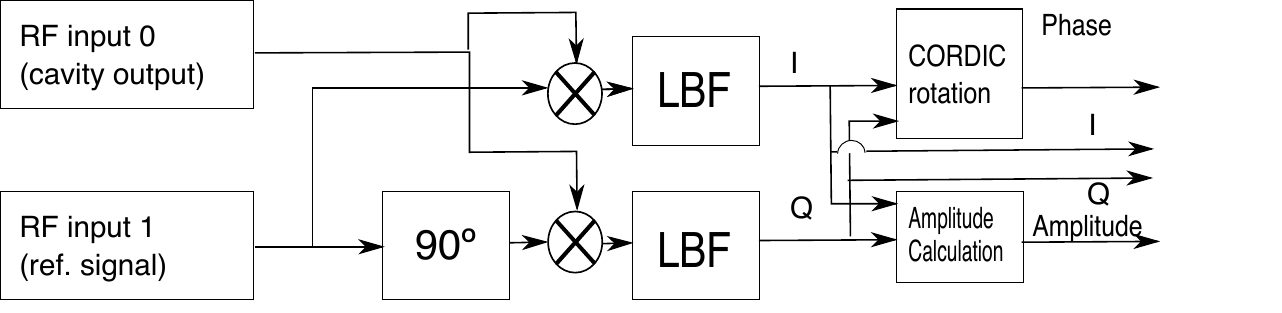}\protect\caption{\label{fig:I/Q-demodulation-FPGA}I/Q demodulation and phase and amplitude
detection in the FPGA }
\end{figure}

The direct demodulation of the acquired RF signal, which is the cavity
output $V_{o}^{cav}$, is performed in the acquisition FlexRIO FPGA
card following the scheme described in Figure \ref{fig:I/Q-demodulation-FPGA}.
Once the I and Q components are calculated, they are independently
controlled following diverse control strategies in order to obtain
corrected values that are related to the desired phase and amplitude
values. This corrected I' and Q' components are generated as system
outputs and used as a modulator baseband inputs, as shown in Figure
\ref{fig:Experimental-setup-description}. The modulator model used
in this work is the ADL5385 from Analog Devices, which takes the reference
RF signal as the local oscillator input and modulates it according
to the calculated I' and Q' values. This modulated signal is directed
to the cavity input, thus closing the feedback loop.

The second loop present in LLRF systems takes care of the variations
in the resonant frequency of the cavity. As main sources of frequency
perturbation in the resonant cavity are much slower than the RF signals,
such as thermal effects or Lorentz force detuning, the frequency tuning
loop can be considered a slow control loop\cite{LLRFlibu}. The actuator
consists of two plungers  attached to two stepper motors respectively
that change the geometry of the cavity in order to keep its resonant
frequency matched to the desired one. The slow loop for the resonance
frequency tuning is implemented measuring the phase variations in
the digital controller.

Those loops require a bandpass filtering of the sampled signals to
be sampled around the $f_{center}$ in order to reduce the error introduced
by the aliasing.

\subsection{Phase and Amplitude feedback control loop implementation}

Working under the PXIe architecture, control structures and signal
processing can be implemented in both the real-time controller and
the FPGA cards. Implementing all the functionalities purely in the
FPGA, faster loop rates can be obtained, leading to higher bandwidths.
The disadvantage of this solution is a more complex and slower design
and prototyping process.

On the other hand, the use of the real-time controller in combination
with the FPGAs, leads to a much more flexible and fast design. But,
with this methodology a significative reduction of the system bandwidth
is obtained.

So, at this point, two models are proposed in order to develop the
control loops of the LLRF system: the first one closing the loop over
the real-time controller, and the second one which uses exclusively
FlexRIO FPGA cards.

\subsection*{Real-time controller based approach}

One of the main motivations of this work, besides the development
of a valid LLRF system that meets the requirements of the projected
heavy-ion accelerator, is to develop a flexible and modular testbench
in which different tests can be carried out. The testbench must be
able to easily modify and add new functionalities as they are needed. 

So, in order to get the desired agility in the design process, the
implementation of the phase and amplitude loop is separated in two
parts: the FPGA part and the real-time controller part.

\begin{figure}
\centering{}\includegraphics[width=0.85\columnwidth]{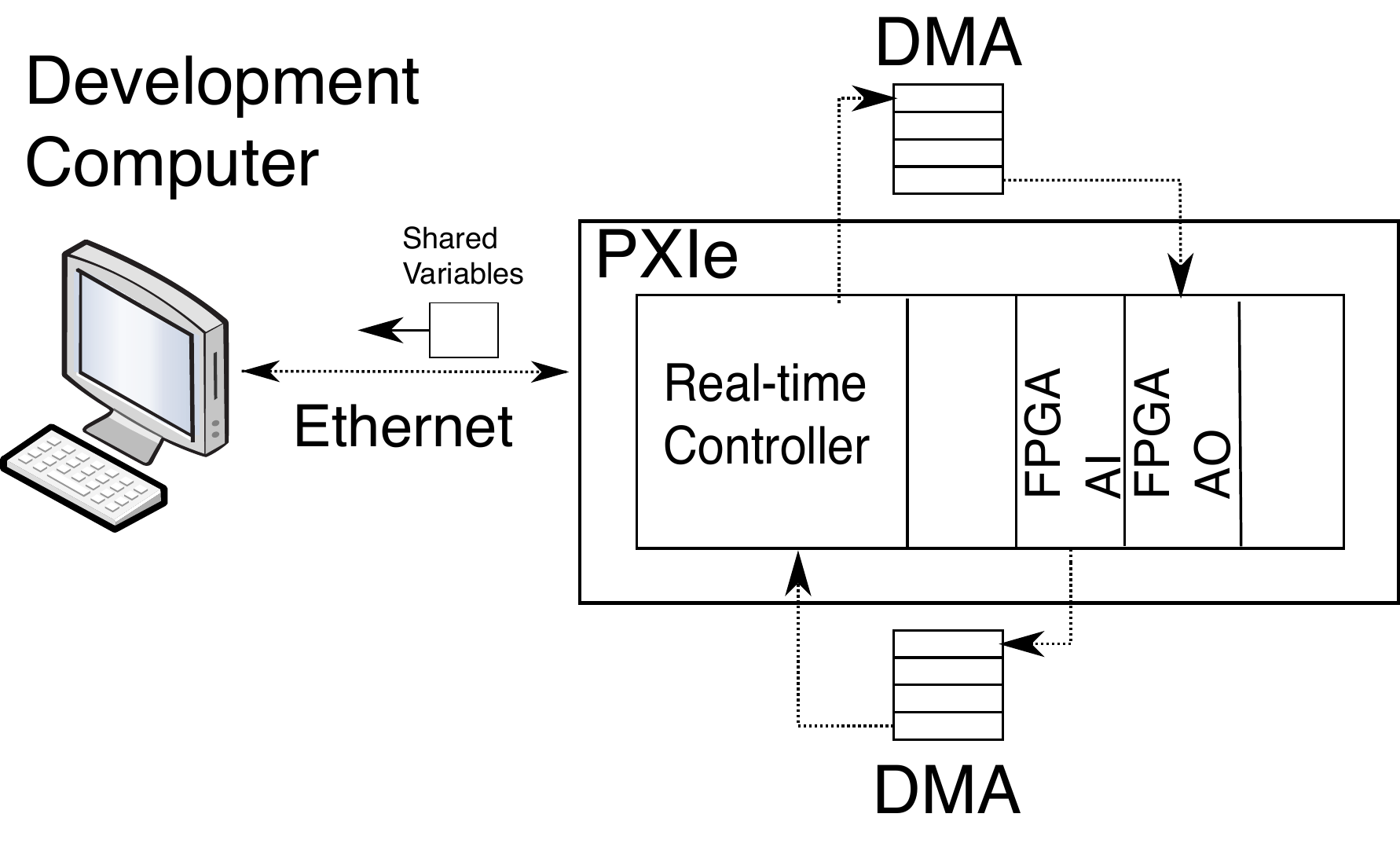}\protect\caption{\label{fig:Schematic-description-of}Schematic description of the
real-time controller based approach }
\end{figure}

This section is focused in the phase and amplitude loop, as it is
the most time-restrictive loop in the LLRF control.

\begin{figure}
\centering{}\includegraphics[width=0.9\columnwidth]{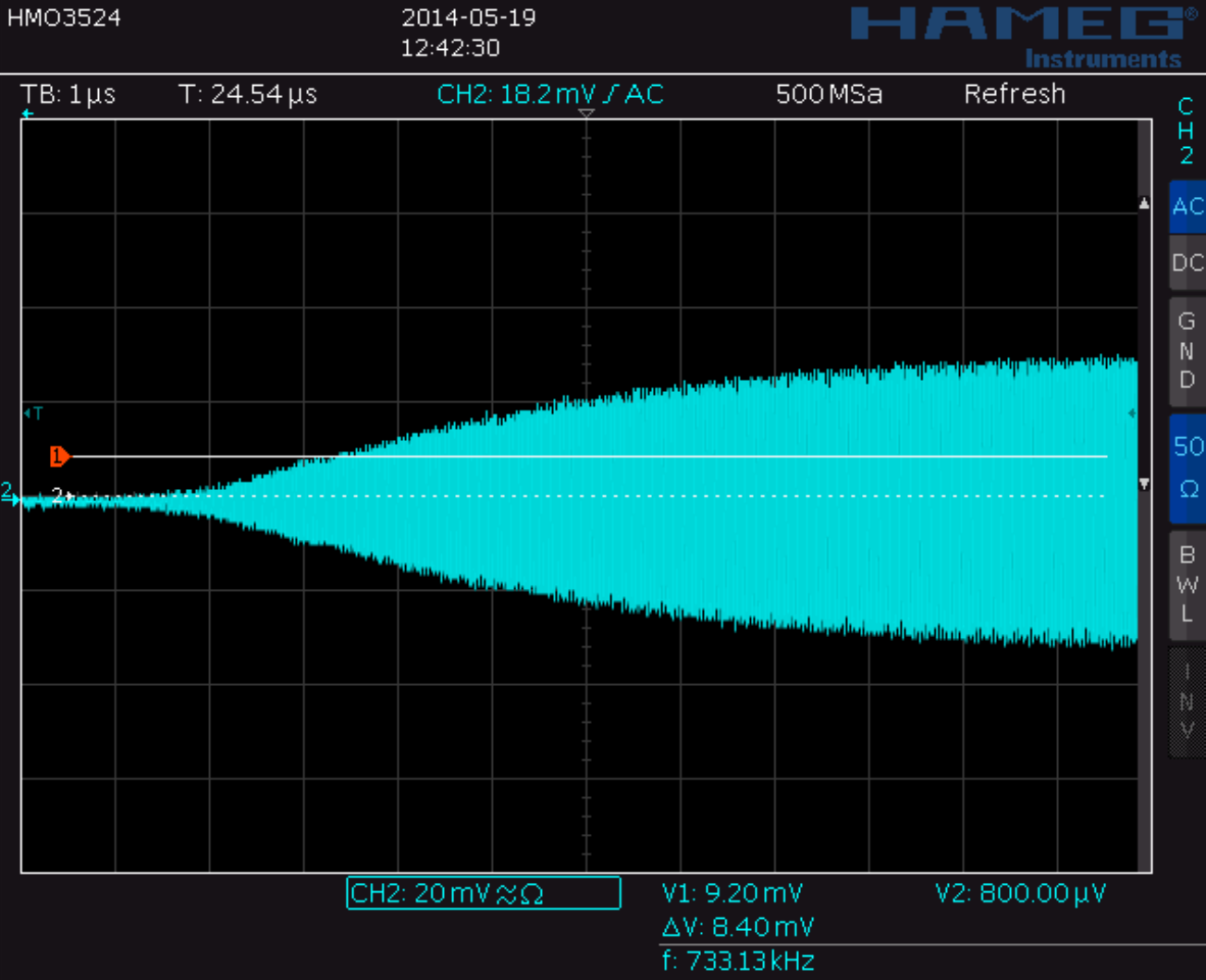}\protect\caption{Cavity output measurement showing cavity filling time \label{fig:Cavity-output-measurement-1}}
\end{figure}

In this approach, three devices are used under the PXIe to close the
phase and amplitude loop. These are:
\begin{itemize}
\item FlexRIO FPGA card which acquires the cavity output $V_{o}^{cav}$
and the reference signal from the RF generator. It also performs time-critical
operations such as I/Q demodulation and phase and amplitude detection.
\item Real-time controller in charge of not so time-critical signal processing,
control actions and monitoring. A more detailed description of the
implemented design is given in the following paragraphs.
\item Multifunction FPGA which is used to generate the corrected I'/Q' values.
\end{itemize}
The communication between the FPGA cards and the real-time controller
is done through Direct Memory Access (DMA), which allows the FPGA
to use the host RAM as if it were its own. This protocol adds an amount
of latency to the feedback loop resulting in a $20KHz$ bandwidth,
that is, it takes a minimum amount of $50\mu s$ to close the loop.
As the cavity filling time has been measured to be about $6ms$ (see
Figure \ref{fig:Cavity-output-measurement-1}), a compensation network,
a lag network more specifically, has been designed and implemented
in order to reduce the plant bandwidth, thus making the cavity response
suitable to be controllable with the given feedback loop speed. This
compensation network is applied separately for both I' and Q' control
signals. 

\begin{figure}
\centering{}\includegraphics[width=0.9\columnwidth]{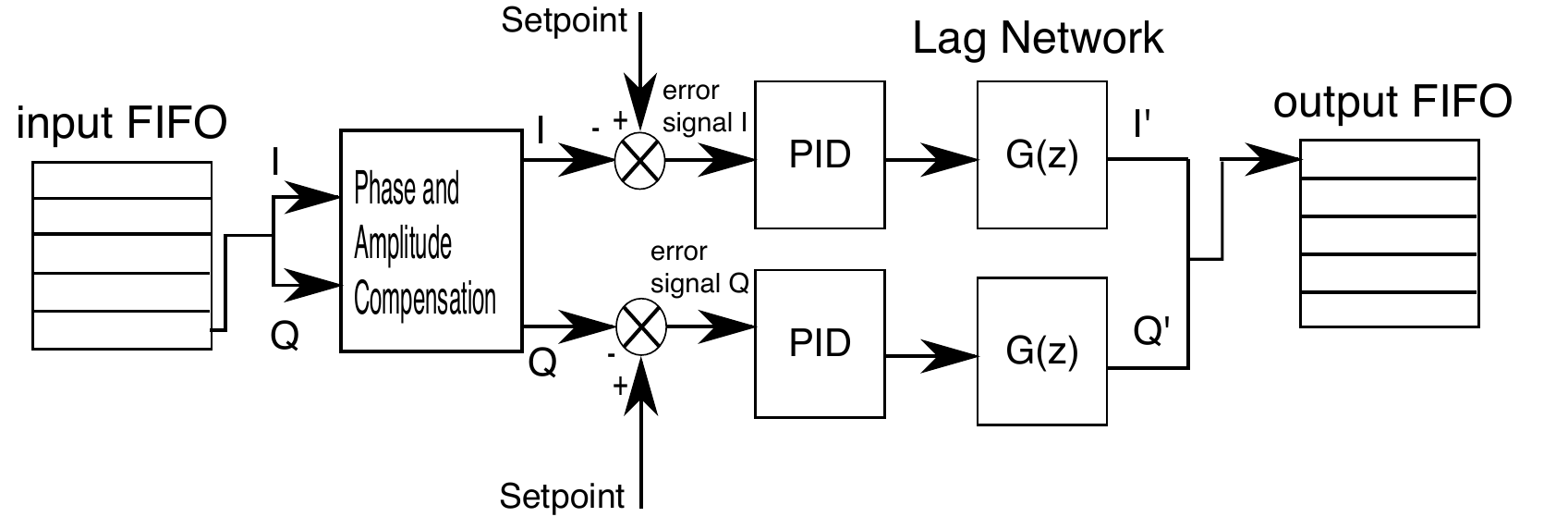}\protect\caption{\label{fig:Implemented-signal-processing}Implemented signal processing
and control structure in the real-time controller}
\end{figure}

In addition, to the lag network, more functionalities are implemented
in the controller. The I/Q components coming from the acquisition
FlexRIO FPGA are compensated to fix the phase difference induced by
the cables present in the experimental set up. Cavity losses are also
compensated before the I/Q components reach the controller. This procedure
is described in Figure \ref{fig:Implemented-signal-processing}.

Finally, in order to minimize unnecessary latency sources, all the
GUI is moved to a regular PC, which is connected in a LAN with the
PXIe via Ethernet as Host controller. All the necessary data to monitor
the relevant data of the phase and amplitude loop as control and error
signals are sent from the PXIe to the host PC using LabVIEW network
shared variables as a low priority task.

This approach allows the fast testing of different control strategies
before moving to a pure FPGA based implementation.

\subsection*{Pure FPGA approach}

The second approach is focused in obtaining the highest possible throughput.
This is more oriented to a final implementation of a previously tested
configuration.

\begin{figure}
\centering{}\includegraphics[width=0.85\columnwidth]{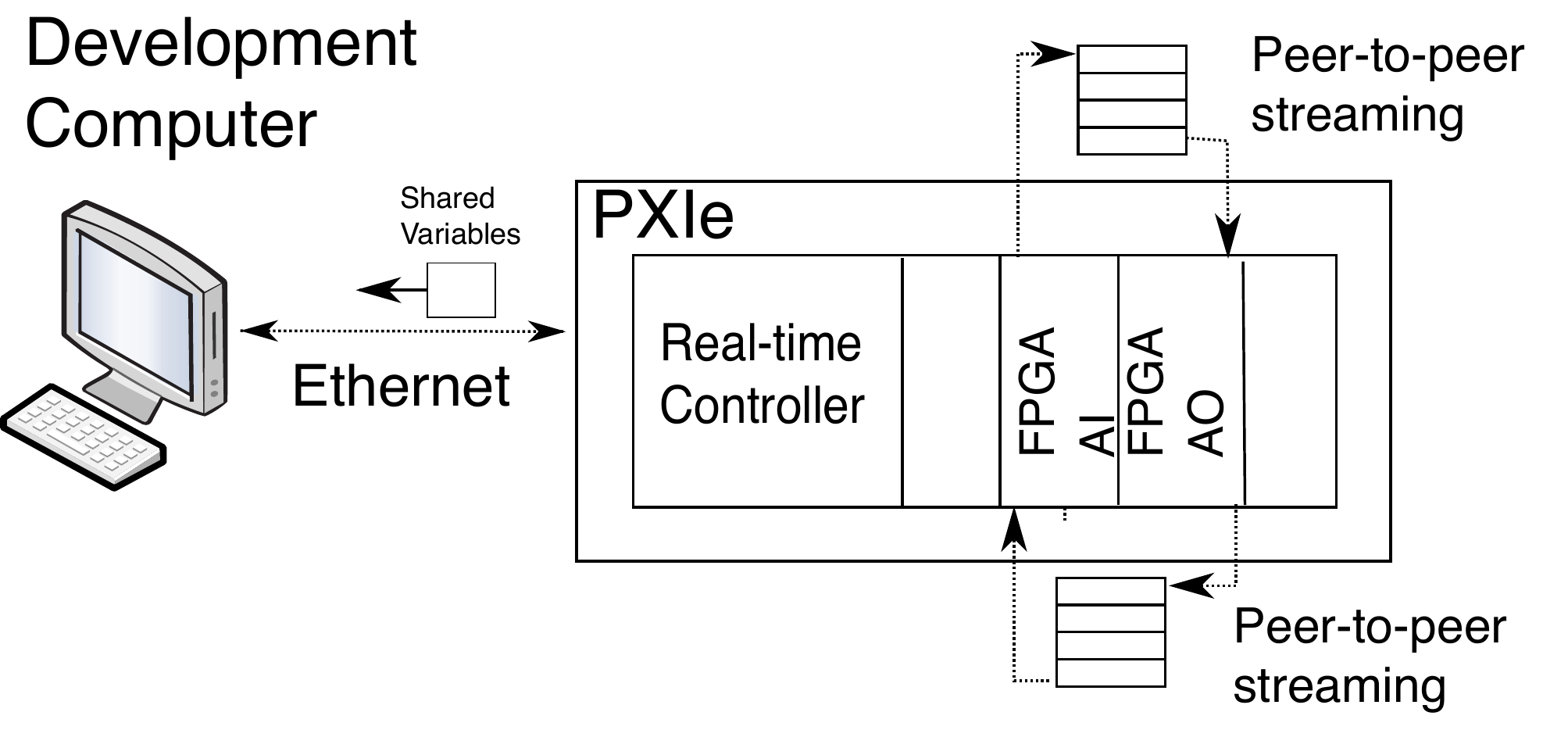}\protect\caption{\label{fig:fpga approach}Schematic description of the approach based
exclusively in FPGAs}
\end{figure}

In this case, the control loop is closed over two FlexRIO FPGA cards,
as shown in Figure \ref{fig:fpga approach}.

Using two FlexRIO cards inside a single PXIe chassis allows to use
peer-to-peer streaming to communicate them. NI peer-to-peer (P2P)
streaming technology uses PCI Express to enable direct, point-to-point
transfers between multiple instruments without sending data through
the host processor or memory. This enables devices in a system to
share information without burdening other system resources \cite{p2p}.
Because the chassis backplane switches provide direct links to the
slots occupied by the FlexRIO cards, there is no need to transfer
data through the host controller or use system resources such as the
CPU or host memory. As a result, high speed transfers are obtained
between FPGA cards, in the order of some hundred MHz. So taking advantage
of the P2P streaming, the phase and amplitude loop bandwidth is not
limited by the data transfer rates between devices as happens in the
previous approach.

Similarly to the first methodology, a FlexRIO card handles the data
acquisition and I/Q demodulation and the other one generates the corrected
I'/Q' components. The signal processing and control architecture described
in Figure \ref{fig:Implemented-signal-processing} can be implemented
in any of them. However, a good practice is to split this code into
both FPGAs to avoid running out of resources.

In this case, the design process is more complex due to the limited
LabVIEW toolkits for FPGA and also for the restriction in data types
inherent to FPGAs. Besides the complexity, there is also the drawback
of longer development times caused by compilation processes. In consequence,
the use of this approach is most oriented to final implementations,
after a successful testing stage.

\subsection{Frequency tuning loop}

Next, the frequency loop has been designed and implemented. For this
loop, time constraints are less restrictive, since the dynamic of
the perturbances is much more slower than the RF signal \cite{LL}.
In this case, the feedback loop rate obtained closing the loop over
the real-time controller is enough to keep the resonant frequency
fixed against external disturbances, even without adding compensation
networks. So, the signal processing and control architecture of the
tuning loop is implemented in the RT controller.

When the cavity is on resonance, the phase difference between the
input and the output has been measured to be -108º in the current
experimental setup (see Figure \ref{fig:Frequency-response-of}).
The frequency loop keeps this value stable against perturbations in
order to minimize the reflected power in the cavity input. The phase
difference is calculated on the FPGA following the scheme described
in Figure \ref{fig:I/Q-demodulation-FPGA}. This value is transferred
to the real-time controller through DMA and corrected using a PI controller
to keep the measured phase matched to the resonance setpoint. The
obtained control signal acts on two stepper motors that move two pistons
respectively which change the cavity geometry.

\section{Experimental results}

In this section, the results related to the two loops of the LLRF
control are presented. The initial tests have been performed following
the first methodology introduced in Section 3.1 defined as real-time
controller based approach.

Focusing in the phase and amplitude loop, the first step has been
the implementation of the compensation network. Lag compensators can
be used to adjust the frequency response of a system \cite{lagnetwork}.
This compensation network can be used to provide better stability,
better performance and general improvement. 

In this particular case, the goal is to slow down the plant dynamic
in order to make the closed-loop bandwidth suitable. The first order
compensator is designed to change the cavity filling time from its
nominal value of 6$\mu s$ (see Figure \ref{fig:Cavity-output-measurement-1})
up to some hundreds of $\mu s$. The compensator in its discrete form
is:

\begin{equation}
H_{lag}(z)=\frac{0.002}{z-0,998}
\end{equation}

\begin{figure}
\centering{}\includegraphics[width=0.9\columnwidth]{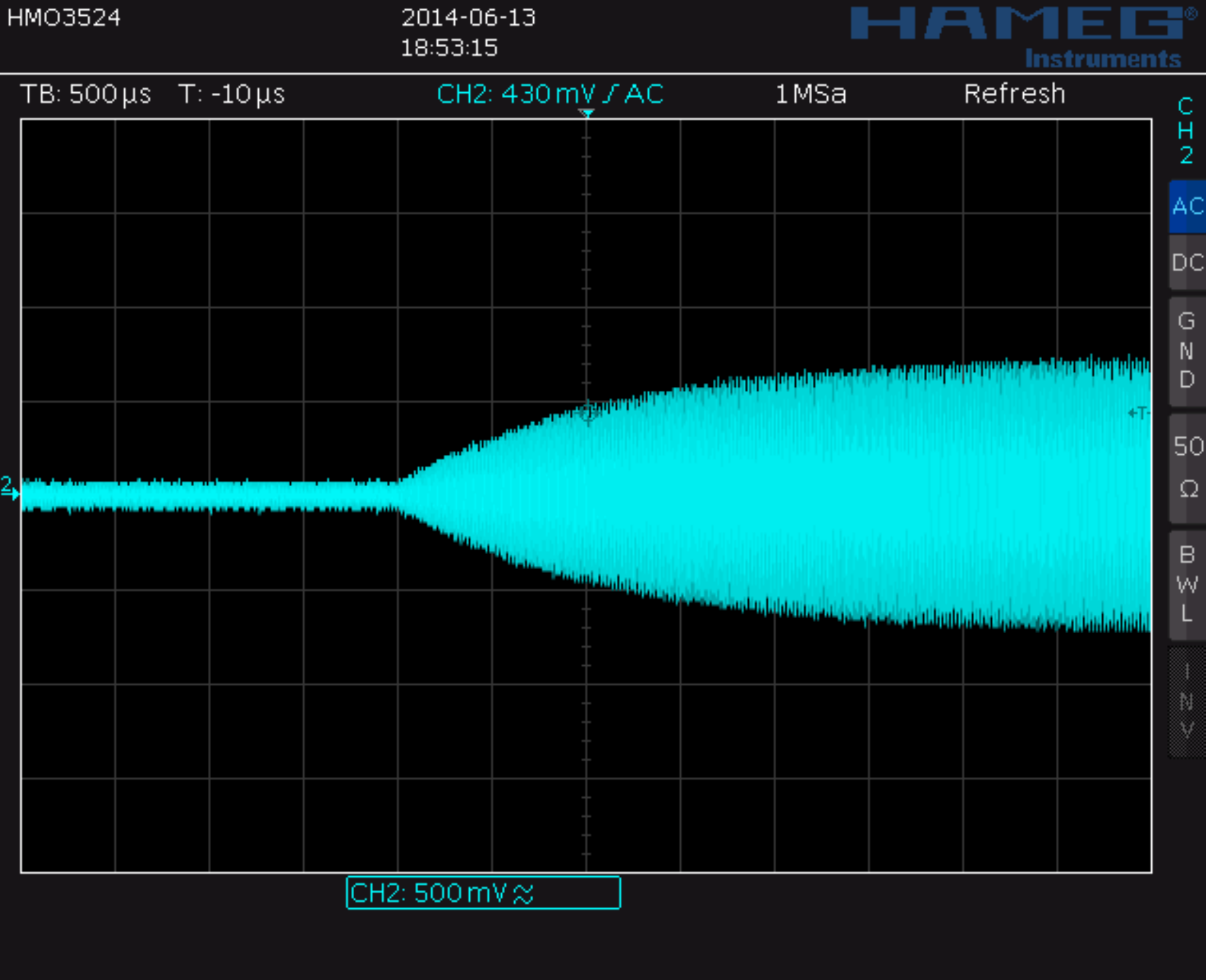}\protect\caption{\label{fig:Cavity-filling-time}Cavity filling time with lag compensation}
\end{figure}

Applying this dynamics, the measurement of the cavity output validates
the implemented lag compensation as seen in Figure \ref{fig:Cavity-filling-time},
where a 1.5 ms filling time can be observed. Simply adjusting the
location of the zeroes and poles of the implemented $H_{lag}(z)$
filter, the frequency response of the plant can be modified to change
the system bandwidth and allow to test different control strategies.

\begin{figure}
\centering{}\includegraphics[width=0.9\columnwidth]{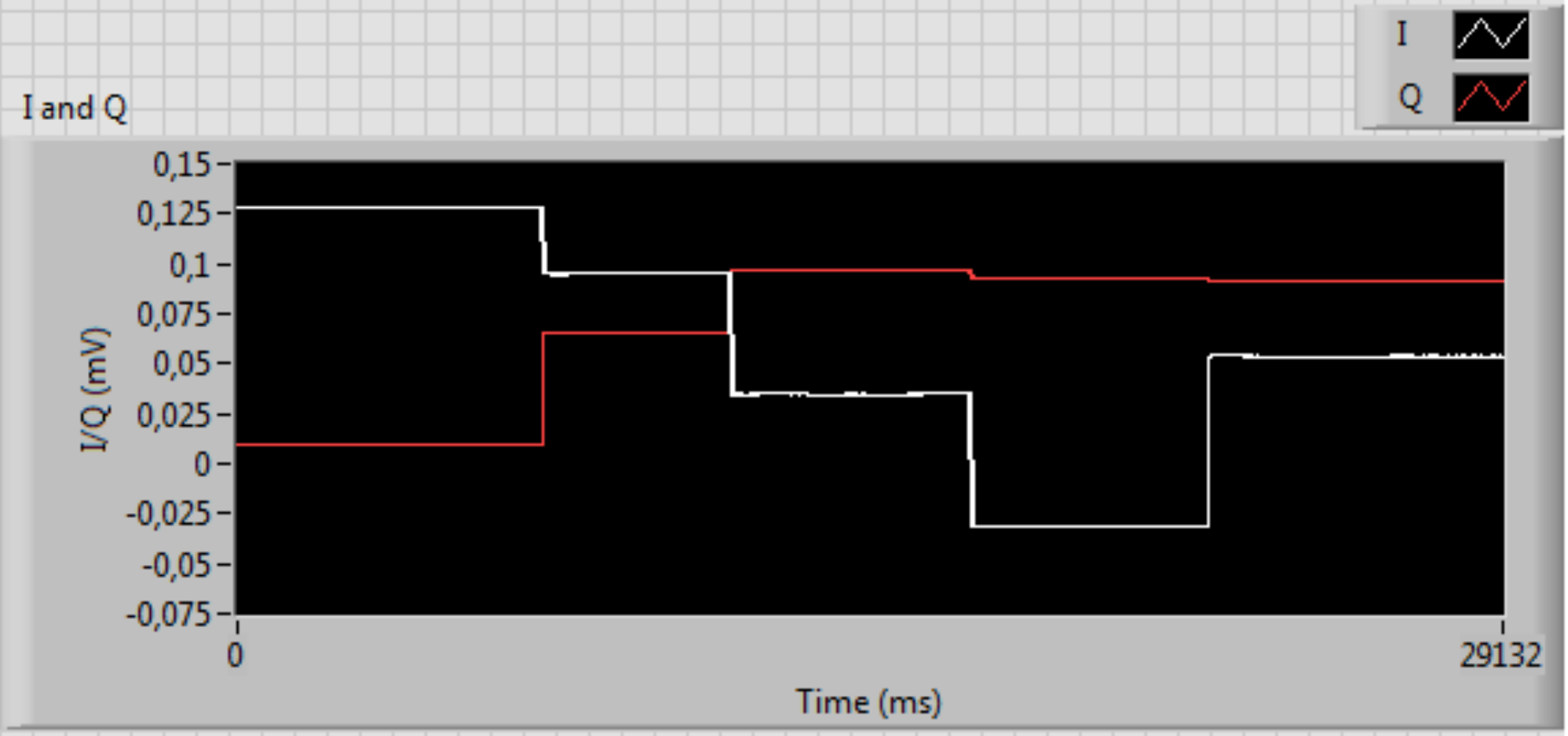}\\
\includegraphics[width=0.9\columnwidth]{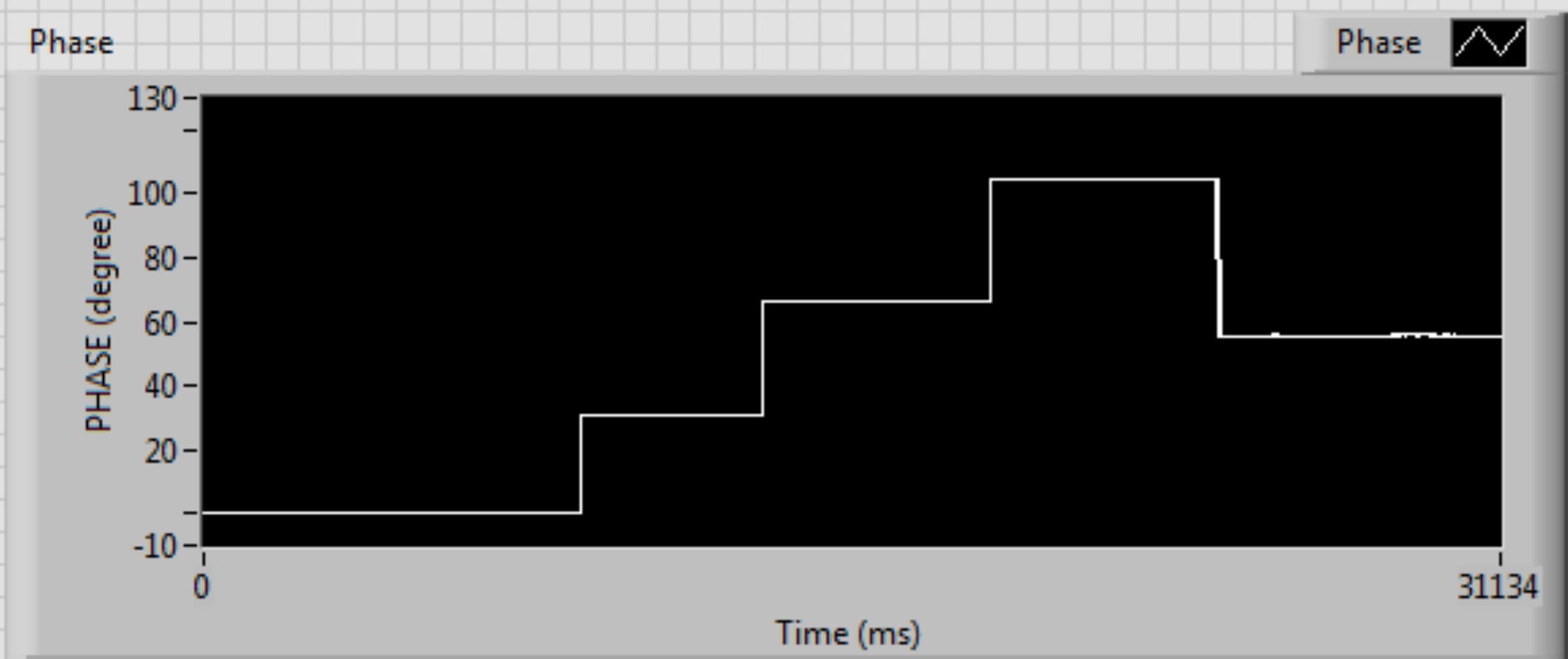}\protect\caption{\label{fig:I-and-Q reult}I and Q components measured at cavity output
(up) against arbitrary phase reference changes and its respective
measured phase (down)}
\end{figure}

\begin{figure}
\centering{}\includegraphics[width=0.9\columnwidth]{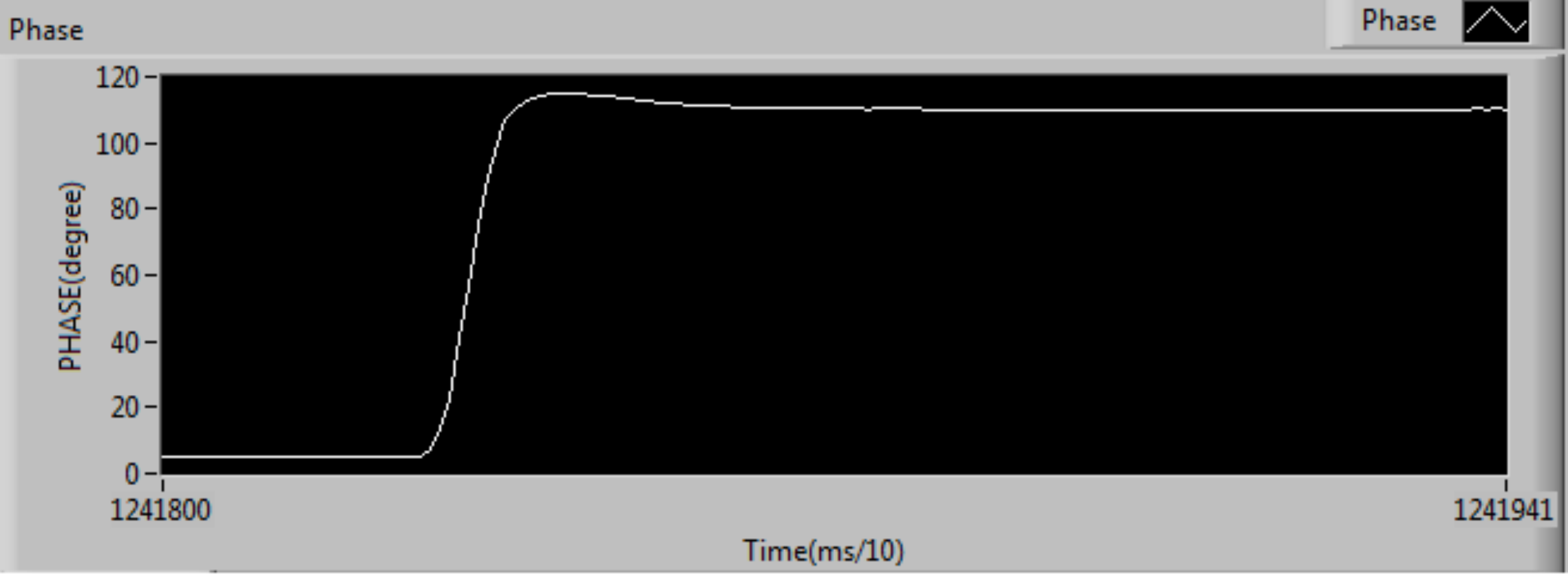}\protect\caption{\label{fig:Step-response-of}Step response of the the phase }
\end{figure}

Using this implementation, phase and amplitude control tests have
been carried out. The phase and amplitude references are externally
set and the result is monitored. In Figure \ref{fig:I-and-Q reult}
is shown the obtained I/Q and phase response. A more detailed time
response of the phase is presented in Figure \ref{fig:Step-response-of}.
The system response against a reference change shows low overshoot
and error in steady condition. The settling time depends on the adjusted
bandwidth.

\begin{figure}
\begin{centering}
\includegraphics[width=0.9\columnwidth]{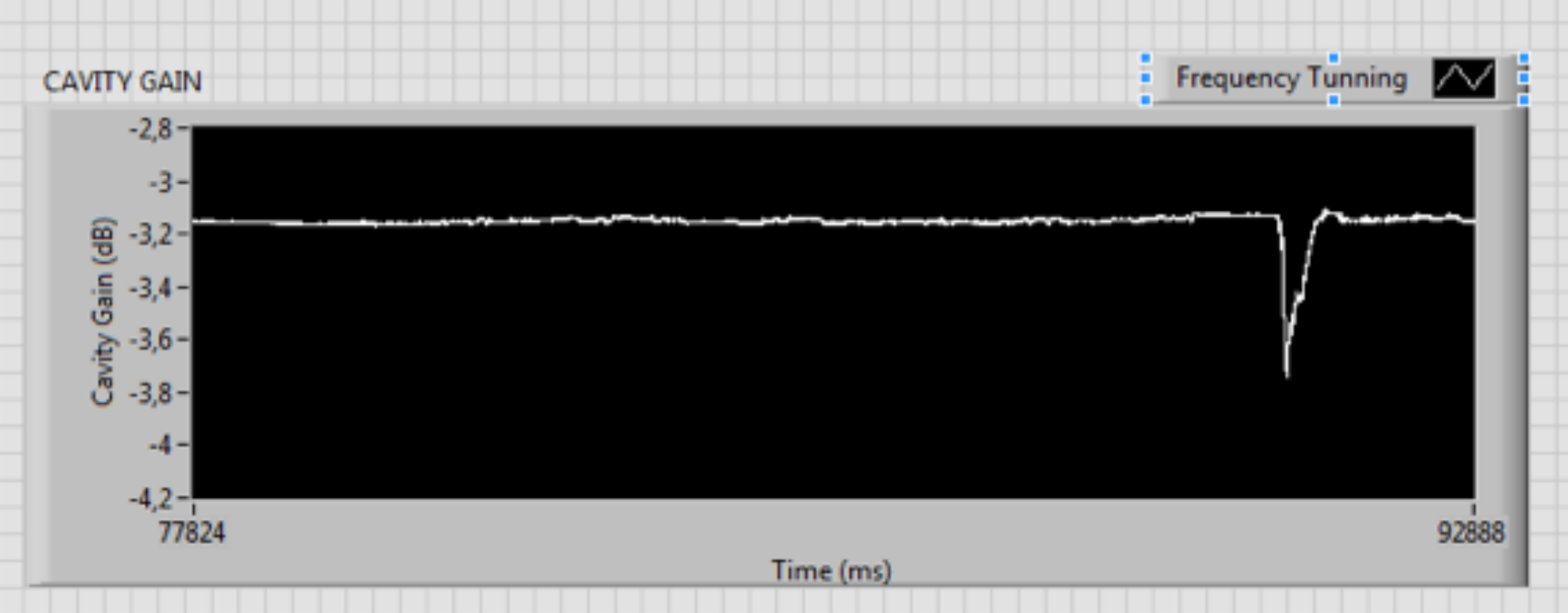}
\par\end{centering}

\centering{}\protect\caption{Resonant frequency compensation against perturbation\label{fig:Resonant-frequency-compensation-1-1}}
\end{figure}

Regarding the results of the tunning loop, the Figure \ref{fig:Resonant-frequency-compensation-1-1}
presents the power gain between the input and the output of the cavity.
It can be observed how the power gain decreases from -3.2dB (in resonance)
to -3.8db in the instant in which a external deforming mechanical
pressure is applied on the top of the cavity and how the stepper motors
act to keep the resonant frequency matched to the nominal one. The
transient response is under 1s in the current configuration.

\section{Summary and conclusions}

This work presents a LLRF architecture for heavy ion linear acceleration
following the specifications of the future LFR in Huelva, Spain. It
is based on a PXIe platform and constitutes a complete test bench
for experimental validation of the prototype. The LLRF system is mainly
digital, offering several advantages over classic analog systems,
such as high flexibility, versatility and reconfigurability. Two different
configurations are presented. The first one combines a realtime controller
with two FPGA cards, leading to a very flexible system but reduced
bandwidth. The second one is only based in FPGA cards improving the
bandwidth but increasing the development cost. The key advantage of
the presented testbench and the two strategies is its flexibility
to add new functionalities in an easy way. 

On the other hand, a subsampling discretization technique is proposed
in order to develop a fast data acquisition system avoiding excessive
equipment costs. The main drawback when using subsampling, the increment
of the clock jitter, is not excessive in this case due to the relatively
small difference between the original and final frequencies.

Initial results validate the proposed scheme. A lag compensator ajusts
the system bandwidth. allowing the use of the two proposed strategies.
Two control loops are presented: the phase and amplitude loop based
in IQ modulation and demodulation and the frequency tuning loop. The
controllers are implemented in the RT system and the FPGA card, taking
advantage of the tools offered by a LabVIEW based environment.

\section{Acknowledgments}

The authors are very grateful by the partial support of this work
to the Basque Goverment by mean of the projects Grupos de Investigación
GIU06/04 and GIU08/01. 

\bibliographystyle{IEEEtran}
\bibliography{bib}

\end{document}